\documentclass[cameraready]{Interspeech}

\title{Nudging Hidden States: Training-Free Model Steering for Chain-of-Thought Reasoning in Large Audio-Language Models}

\author[affiliation={}, equalcontribution]{Lok-Lam}{Ieong}
\author[affiliation={}, equalcontribution]{Chia-Chien}{Chen}
\author[affiliation={}, equalcontribution]{Chih-Kai}{Yang}
\author[affiliation={}]{Yu-Han}{Huang$^{\dagger}$}
\author[affiliation={}]{An-Yu}{Cheng$^{\dagger}$}
\author[affiliation={}]{Hung-yi}{Lee}

\address{
  National Taiwan University, Taiwan
}
\email{chihkaiyang1124@gmail.com, hungyilee@ntu.edu.tw}


\newcommand{\red}[1]{\textcolor{red}{#1}}
\newcommand{\green}[1]{\textcolor{green!45!black}{#1}}
\usepackage{comment}
\usepackage{cite}
\usepackage{hyperref}
\usepackage{url}
\usepackage{makecell}
\usepackage{multirow}
\usepackage{xcolor}
\usepackage{relsize}
\usepackage{graphicx}
\usepackage{subcaption}
\usepackage{caption}

\begin{document}

\maketitle

\begin{abstract}
  Chain-of-thought (CoT) prompting has been extended to large audio-language models (LALMs) to elicit reasoning, yet enhancing its effectiveness without training remains challenging. We study inference-time model steering as a training-free approach to improve LALM reasoning. We introduce three strategies using diverse information sources and evaluate them across four LALMs and four benchmarks. Results show general accuracy gains up to 4.4\% over CoT prompting. Notably, we identify a cross-modal transfer where steering vectors derived from few text samples effectively guide speech-based reasoning, demonstrating high data efficiency. We also examine hyperparameter sensitivity to understand the robustness of these approaches. Our findings position model steering as a practical direction for strengthening LALM reasoning.
\end{abstract}

\section{Introduction}

Large audio-language models (LALMs)~\cite{voxtral, phi4, qwen25omni, af3, desta25, copilot, tp1, lin2025preliminary}, which extend large language models (LLMs)~\cite{hurst2024gpt, grattafiori2024llama, yang2025qwen3} with auditory understanding~\cite{audiolens, sake}, have recently achieved substantial progress. They demonstrate strong auditory perceptual capabilities~\cite{audiolens, dynamicsuperb} and are increasingly regarded as a promising foundation for universal and interactive auditory intelligence~\cite{arora2025on, yang-etal-2025-towards-holistic, wang2025towards}. However, reasoning remains a fundamental limitation~\cite{mmau, mmaupro, sakura} that prevents these models from fully realizing this vision.

Reasoning has long been central to artificial intelligence. In LLMs, Chain-of-Thought (CoT) prompting~\cite{cot, cot2} is a representative approach for eliciting structured, step-by-step reasoning. Motivated by its success, recent work extends CoT to LALMs~\cite{audiocot}, with further improvements via supervised reasoning data or reinforcement learning~\cite{audioreasoner, audiothinker}. However, these methods require additional supervision and substantial training cost. This raises a key question: \emph{Can we enhance CoT reasoning in LALMs at inference time without extra training?}

Model steering~\cite{subramani-etal-2022-extracting, pai2025billy, turner2025steering, refusal, venhoff2025understanding, zhu2025self, sinii-etal-2025-steering} offers a training-free alternative by manipulating hidden states. In LLMs, steering vectors have been used for style control~\cite{subramani-etal-2022-extracting, pai2025billy}, safety alignment~\cite{refusal}, and performance improvement~\cite{zhu2025self, sinii-etal-2025-steering}. In LALMs, steering has been applied to attribute recognition~\cite{audiolens}, hallucination mitigation~\cite{lin2025adaptive}, and safety alignment~\cite{lin2025sarsteer}, but its potential for enhancing reasoning remains underexplored.

\begin{figure}[t]
    \centering
    \includegraphics[width=0.99\linewidth]{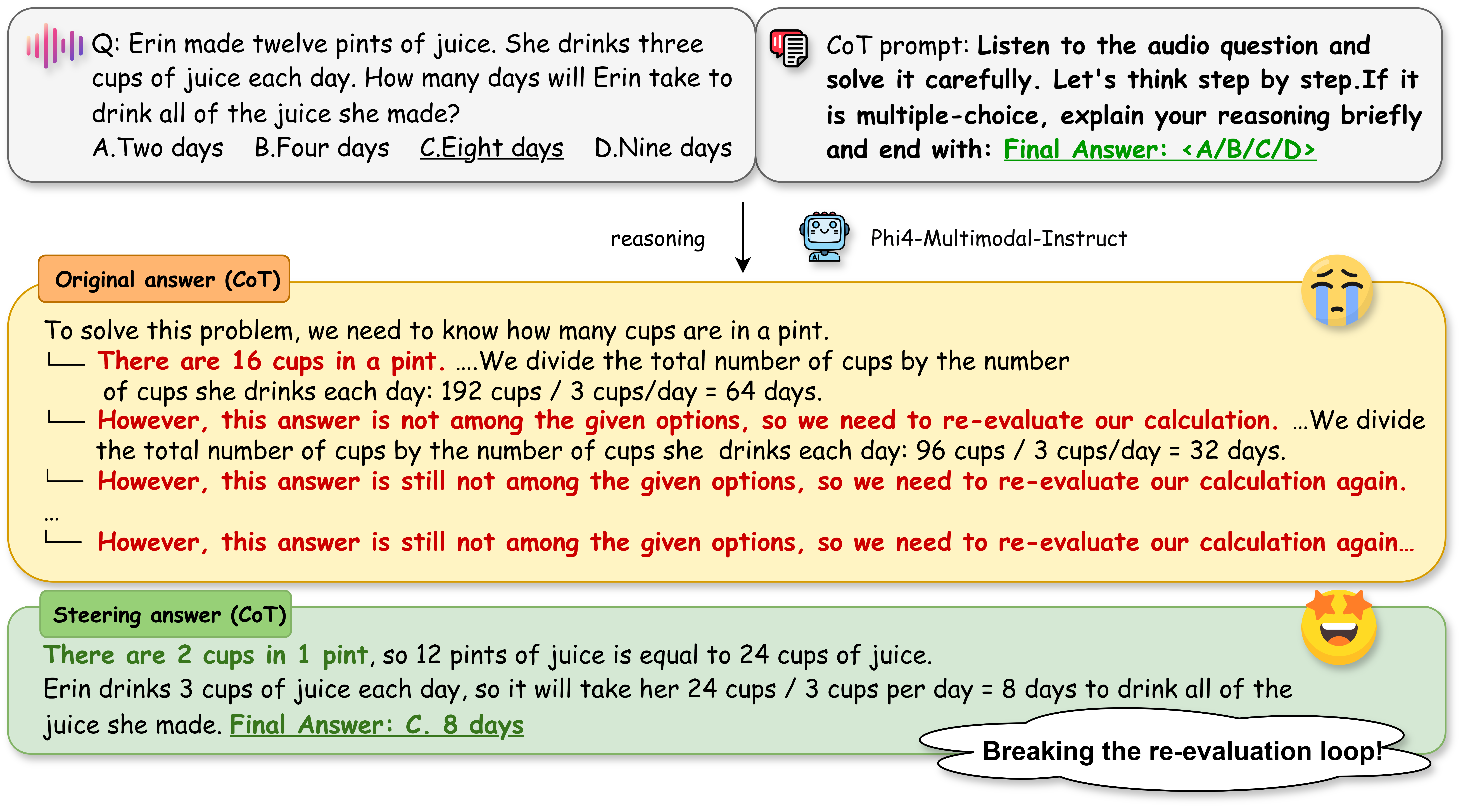}
    \caption{Example of reasoning enhanced by steering.}
    \label{fig:example}
    \vspace{-15pt}
\end{figure}

This work investigates model steering as a representation-level intervention to improve CoT reasoning in LALMs. Reasoning-oriented steering directions are derived from the difference between CoT and non-CoT hidden states and injected during decoding. We propose three variants: \textbf{Vanilla Steering}, which extracts instance-specific vectors; \textbf{Speech-derived Generalized Steering} (SGS), which extracts a shared vector from auxiliary spoken data; and \textbf{Text-derived Generalized Steering} (TGS), which derives steering directions from text-only data and transfers them to speech-based reasoning. Figure~\ref{fig:example} shows a qualitative example where steering improves intermediate reasoning and corrects the final prediction.

Extensive experiments on four advanced LALMs and four speech-based benchmarks show that steering generally improves CoT performance, with up to 4.4\% absolute accuracy gains over CoT. Vanilla steering surpasses self-consistency~\cite{wang2023selfconsistency} under a comparable computational budget while requiring fewer decoding operations. SGS and TGS demonstrate that effective steering vectors can be extracted without instance-specific access, achieving competitive improvements across models. Notably, TGS achieves higher average accuracy than CoT across all models despite deriving steering vectors purely from text data. Hyperparameter sensitivity and data-efficiency analyses further show that generalized steering methods are more stable than instance-specific steering, with TGS requiring fewer samples to reach competitive performance, highlighting its stability and data efficiency for enhancing speech-based reasoning.

In summary, this work (1) introduces a training-free steering framework for enhancing CoT reasoning in LALMs, (2) demonstrates its effectiveness and computational efficiency across multiple models and benchmarks, (3) reveals the feasibility of generalized and cross-modal steering directions, and (4) provides empirical insights into the stability and data-efficiency characteristics of steering-based interventions.

\begin{figure*}[!t]
  \centering
  \includegraphics[width=0.99\textwidth]{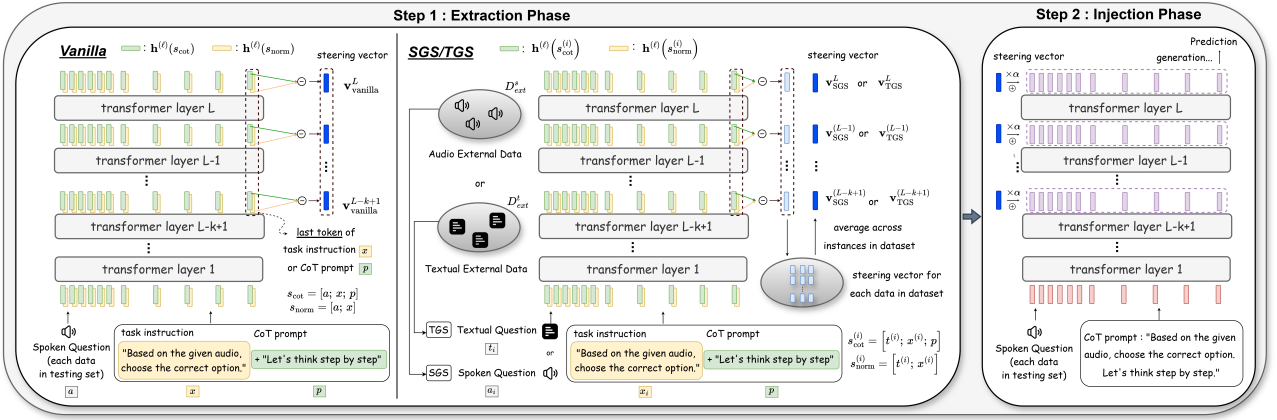}
  
  \caption{Overview of our three proposed methods in the extraction phase, along with the subsequent injection phase.}
  \label{fig:overview}
\end{figure*}

\section{Methodology}

Model steering consists of two phases: (1) an \emph{extraction phase}, where steering vectors are derived, and (2) an \emph{injection phase}, where the vectors are applied during generation. We describe these phases and present three extraction methods (Figure~\ref{fig:overview}).

\subsection{Extraction Phase}

Given a model with $L$ layers, the extraction phase constructs steering vectors from hidden states $\mathbf{h}_t^{(\ell)}(s)\in\mathbb{R}^{d}$ at layer $\ell\in[1,L]$ and token position $t$ for input $s$. Following prior work, we extract steering directions from the last $k$ layers~\cite{valentino2025mitigating, huang2025mitigating, wang-etal-2025-improving-llm, tang-etal-2025-enhancing} at the final prompt token~\cite{refusal, stolfo2025improving, braun2025understanding, rimsky-etal-2024-steering}, denoted by $\bar{\mathbf{h}}^{(\ell)}(s)$ for $\ell \in [L-k+1, L]$. Below, we introduce three extraction strategies.


\subsubsection{Vanilla Steering}


In \emph{Vanilla Steering}, the steering vector is constructed dynamically for each test sample at inference time. For a given sample, let $a$ denote the audio input, $x$ the task instruction, and $p$ a fixed short chain-of-thought cue. We form two inputs:
\begin{equation}
\begin{aligned}
s_{\text{cot}}  &= [a;\,x;\,p], \qquad
s_{\text{norm}} = [a;\,x].
\end{aligned}
\label{eq:format}
\end{equation}%
For each selected layer $\ell$, the steering vector is defined as
\begin{equation}
\mathbf{v}_{\text{vanilla}}^{(\ell)} = \bar{\mathbf{h}}^{(\ell)}(s_{\text{cot}}) - \bar{\mathbf{h}}^{(\ell)}(s_{\text{norm}}).
\end{equation}
Since $\mathbf{v}_{\text{vanilla}}^{(\ell)}$ depends only on the hidden states under different prompting conditions for the same input, no ground truth or external supervision are involved. The procedure therefore constitutes a valid training-free inference-time intervention.

\subsubsection{Speech-derived Generalized Steering (SGS)}

A limitation of Vanilla Steering is its computational overhead: extracting a sample-specific steering vector requires additional forward passes for each test input. This motivates constructing a shared steering direction that can be reused across samples.

To this end, we propose \emph{Speech-derived Generalized Steering} (SGS). An external auxiliary spoken dataset $\mathcal{D}^{s}_{\text{ext}}$ is used to compute a shared steering vector applied uniformly to all test samples. For each example $i \in \mathcal{D}^{s}_{\text{ext}}$, let $a^{(i)}$ denote the audio input and $x^{(i)}$ the task instruction. We construct
\begin{equation}
s^{(i)}_{\text{cot}}=[a^{(i)};\,x^{(i)};\,p], \qquad s^{(i)}_{\text{norm}}=[a^{(i)};\,x^{(i)}].
\end{equation}
Following the Difference-in-Means paradigm~\cite{diff-in-means, marks2024the, tigges2023linear}, we average the differences over $\mathcal{D}^{s}_{\text{ext}}$ to obtain a shared steering vector:
\begin{equation}
\mathbf{v}_{\text{SGS}}^{(\ell)}
=
\frac{1}{|\mathcal{D}^{s}_{\text{ext}}|}
\sum_{i\in\mathcal{D}^{s}_{\text{ext}}}
\Big(
\bar{\mathbf{h}}^{(\ell)}(s_{\text{cot}}^{(i)})
-
\bar{\mathbf{h}}^{(\ell)}(s_{\text{norm}}^{(i)})
\Big).
\end{equation}
Unlike Vanilla Steering, $\mathbf{v}_{\text{SGS}}^{(\ell)}$ is computed once and reused across test sets. We evaluate whether such a shared direction can serve as a general reasoning-oriented steering signal.

\subsubsection{Text-derived Generalized Steering (TGS)}



While SGS relies on spoken data for extraction, such data may be less accessible than text in practice. We therefore introduce \emph{Text-derived Generalized Steering} (TGS), which derives a shared steering direction from text-only data and examines whether it can transfer to spoken reasoning tasks.

Using an external textual dataset $\mathcal{D}^{t}_{\text{ext}}$, 
for each example $i \in \mathcal{D}^{t}_{\text{ext}}$, 
let $t^{(i)}$ and $x^{(i)}$ denote the textual input and instruction, respectively. 
We construct
\begin{equation}
s_{\text{cot}}^{(i)}=[t^{(i)};\,x^{(i)};\,p], 
\qquad 
s_{\text{norm}}^{(i)}=[t^{(i)};\,x^{(i)}].
\end{equation}

The shared steering vector is then computed in the same manner as SGS, by applying the Difference-in-Means over $\mathcal{D}^{t}_{\text{ext}}$:
\begin{equation}
\mathbf{v}_{\text{TGS}}^{(\ell)}
=
\frac{1}{|\mathcal{D}^{t}_{\text{ext}}|}
\sum_{i\in\mathcal{D}^{t}_{\text{ext}}}
\Big(
\bar{\mathbf{h}}^{(\ell)}(s_{\text{cot}}^{(i)})
-
\bar{\mathbf{h}}^{(\ell)}(s_{\text{norm}}^{(i)})
\Big).
\end{equation}
The resulting vector is extracted entirely from text-only inputs and then transferred to spoken reasoning tasks at inference time. This setup allows us to assess whether a text-derived steering direction can improve CoT performance in speech-based settings.

\begin{table*}[t]
\centering
\caption{Accuracies (\%) on four evaluation benchmarks and their micro-average (ALL). \textbf{Bold} values denote the best result among the five settings for each model on each benchmark. “Hyperparams.” specifies the last $k$ layers and the scaling factor $\alpha$ used for each configuration. $\Delta$ indicates the difference in average accuracy between our steering methods and the CoT baseline.}
\vspace{-5pt}
\label{tab:main_result}

\setlength{\tabcolsep}{3.6pt}
\renewcommand{\arraystretch}{0.9}
\resizebox{0.98\textwidth}{!}{
\begin{tabular}{ll|c|cccc|cc}
\toprule
Model & Method & Hyperparams. ($k$, $\alpha$) & College ($\uparrow$)  & High School ($\uparrow$) & Elementary ($\uparrow$) & ReveAL-CoT ($\uparrow$) & ALL ($\uparrow$) & $\Delta$ ($\uparrow$) \\
\midrule
\multirow{5}{*}{Voxtral}
& Normal & -- & 29.0  & 36.7 & 56.9 & 53.5 & 47.5 & -- \\
& CoT    & -- & 26.3  & 42.0 & 67.4 & 43.0 & 50.7 & -- \\
\cmidrule{2-9} 
& Vanilla \textbf{(Ours)}           &  5, 0.025 & \textbf{32.0}  & 43.3 & \textbf{68.8} & 56.0 & \textbf{55.0} & \green{\textbf{+4.3}}\\
& SGS \textbf{(Ours)}               & 3, 0.1 & 29.3  & \textbf{43.9} & 65.8 & \textbf{57.0} & 53.8 & \green{\textbf{+3.1}}\\
& TGS \textbf{(Ours)}               & 4, 0.025 & 26.3  & 41.3 & 67.1 & 54.5 & 52.8 & \green{\textbf{+2.1}} \\
\midrule
\multirow{5}{*}{Phi4-mm}
& Normal & -- & 26.0  & 24.8 & 33.1 & 38.5 & 31.1 & -- \\
& CoT    & -- & \textbf{36.0}  & 41.9 & 69.6 & 68.5 & 57.9 & -- \\
\cmidrule{2-9} 
& Vanilla \textbf{(Ours)}           & 5, 0.025 & 35.0  & 47.0 & 69.1 & 69.5 & 59.3 & \green{\textbf{+1.4}} \\
& SGS \textbf{(Ours)}               & 3, 0.05 & 30.0  & 47.0 & 68.8 & \textbf{70.5} & 58.9 & \green{\textbf{+1.0}} \\
& TGS \textbf{(Ours)}               & 3, 0.025 & 30.0  & \textbf{48.9} & \textbf{71.4} & \textbf{70.5} & \textbf{60.4} & \green{\textbf{+2.5}} \\
\midrule
\multirow{5}{*}{Qwen2.5}
& Normal & -- & 40.0  & 45.6 & 52.1 & 51.5 & 48.8 & --\\
& CoT    & -- & \textbf{65.0}  & 79.9 & 83.0 & 71.0 & 77.7 & --\\
\cmidrule{2-9} 
& Vanilla \textbf{(Ours)}           & 3, 0.025 & 56.0  & 79.6 & 83.9 & \textbf{74.0} & 77.6 & \red{\textbf{-0.1}}\\
& SGS \textbf{(Ours)}               & 4, 0.025 & \textbf{65.0}  & 80.7 & \textbf{84.7} & 71.5 & \textbf{78.7} & \green{\textbf{+1.0}}\\
& TGS \textbf{(Ours)}               & 3, 0.05 & 64.0  & \textbf{81.9} & 83.6 & 71.5 & 78.5 & \green{\textbf{+0.8}} \\

\midrule
\multirow{5}{*}{AF3}
& Normal & -- & 26.1  & \textbf{28.2} & 26.0 & 46.5 & 31.0 & --\\
& CoT    & -- & \textbf{27.4}  & 24.2 & 30.4 & 45.5 & 31.5 & --\\
\cmidrule{2-9} 
& Vanilla \textbf{(Ours)}           & 3, 0.1 & 20.0  & 22.2 & 37.6 & \textbf{47.5} & 33.4 & \green{\textbf{+1.9}}\\
& SGS \textbf{(Ours)}               & 1, 0.025 & 24.0  & 21.1 & 36.0 & 42.5 & 31.9 & \green{\textbf{+0.4}}\\
& TGS \textbf{(Ours)}               & 5, 0.1 & 20.0  & 24.1 & \textbf{42.9} & 46.5 & \textbf{35.9} & \green{\textbf{+4.4}} \\
\bottomrule
\end{tabular}
}
\end{table*}
\subsection{Injection Phase}
After obtaining the steering vector $\mathbf{v}^{(\ell)}$ from the extraction phase, we scale it by a coefficient $\alpha$, which controls the steering strength. The scaled vector is injected at inference time into the same set of layers selected during extraction, and the intervention is applied throughout decoding for all token positions.

Let $\mathbf{h}_t^{(\ell)} \in \mathbb{R}^{d}$ denote the original hidden state at layer $\ell$ and token position $t$ during inference. For each selected layer, we modify the hidden state by
\begin{equation}
\tilde{\mathbf{h}}_t^{(\ell)} = \mathbf{h}_t^{(\ell)} + \alpha\, \mathbf{v}^{(\ell)}.
\end{equation}
Following common practice in steering-based interventions, we apply norm-preserving injection to improve stability by rescaling the modified hidden state to match the original $\ell_2$ norm:
\begin{equation}
\hat{\mathbf{h}}_t^{(\ell)}
= \tilde{\mathbf{h}}_t^{(\ell)} \cdot
\frac{\lVert \mathbf{h}_t^{(\ell)} \rVert_2}{\lVert \tilde{\mathbf{h}}_t^{(\ell)} \rVert_2}.
\end{equation}
During inference, $\hat{\mathbf{h}}_t^{(\ell)}$ replaces $\mathbf{h}_t^{(\ell)}$ at all token positions for the selected layers, and the model proceeds with standard forward computation and decoding using the modified states.

LALMs may not always reliably follow CoT instructions, as their instruction-following ability can be weaker after multimodal training~\cite{speechifeval}. Consequently, CoT prompts may not fully induce structured reasoning during decoding. By injecting a steering vector derived from CoT-induced state differences, we reinforce CoT-related activations in the hidden states, encouraging the model to produce more structured reasoning.


\section{Experimental Setups}
\subsection{Models}

We conducted our study in four advanced LALMs: Voxtral-mini-3B (Voxtral)~\cite{voxtral}, Phi4-Multimodal-Instruct (Phi4-mm)~\cite{phi4}, Qwen2.5-Omni-7B (Qwen2.5)~\cite{qwen25omni}, and Audio Flamingo 3 (AF3)~\cite{af3}. Unless otherwise specified, greedy decoding is used for all models.

\subsection{Baselines}

We compare our steering methods with several baselines. \emph{Normal} denotes the default performance of the LALMs, while \emph{CoT}~\cite{cot, cot2} refers to direct chain-of-thought prompting.

We additionally include \emph{self-consistency}~\cite{wang2023selfconsistency}, where each model generates three outputs with temperature $0.5$ and aggregates them via majority voting. This baseline approximates the computational cost of Vanilla Steering, which also requires three forward passes per instance. However, self-consistency remains more expensive because all passes involve full generation, whereas the extraction phase in Vanilla Steering does not.

\subsection{Datasets and Evaluation Benchmarks}

We use BeyondAIME~\cite{beyondaime} (100 samples) as the external dataset for SGS and TGS, strictly disjoint from all evaluation benchmarks. For SGS, math questions are verbalized and synthesized with IndexTTS2~\cite{zhou2025indextts2} to construct $\mathcal{D}^{s}_{\text{ext}}$ (with manual quality checks). For TGS, the original BeyondAIME forms $\mathcal{D}^{t}_{\text{ext}}$.

We tune the number of steered last layers $k$ and the scaling factor $\alpha$ on the spoken GSM8K benchmark~\cite{gsm8k} from SpeechR~\cite{speechr}, used as a development set disjoint from all evaluation benchmarks. We search over $\alpha \in [0.025, 0.2]$ and $k \in {1, \dots, 5}$, and report test performance using the configuration that achieves the best development accuracy.

We evaluate steering on four spoken reasoning benchmarks: \emph{College}, \emph{High School}, and \emph{Elementary} Mathematics from VoxEval~\cite{voxeval}, covering math problems of varying difficulty, and \emph{ReveAL-CoT} from SpeechR~\cite{yang2025speechr}, targeting spoken scientific reasoning. We follow their official evaluation protocols.

\begin{table}[t]
\centering
\caption{Self-consistency (Self-Con) vs. vanilla steering on College (Col), High School (HS), Elementary (Elem) math, and ReveAL-CoT (RC), with overall accuracy (ALL).}
\label{tab:main_result_sc}

\small
\setlength{\tabcolsep}{5.2pt}
\renewcommand{\arraystretch}{0.9}
\resizebox{0.95\linewidth}{!}{
\begin{tabular}{ll|cccc|c}
\toprule
Model & Method & Col & HS & Elem & RC & ALL \\
\midrule
\multirow{2}{*}{Voxtral}
& Self-Con    & 25.0  & 36.3 & 66.9 & 51.0 & 50.4 \\
& Vanilla & \textbf{32.0}  & \textbf{43.3} & \textbf{68.8} & \textbf{56.0} & \textbf{55.0} \\
\midrule
\multirow{2}{*}{Phi4-mm}
& Self-Con    & \textbf{43.0}  & 45.2 & 65.6 & 65.0 & 57.3 \\
& Vanilla & 35.0  & \textbf{47.0} & \textbf{69.1} & \textbf{69.5} & \textbf{59.3} \\
\midrule
\multirow{2}{*}{Qwen2.5}
& Self-Con    & \textbf{58.0}  & 74.1 & 81.8 & 66.5 & 73.8 \\
& Vanilla & 56.0  & \textbf{79.6} & \textbf{83.9} & \textbf{74.0} & \textbf{77.6} \\

\midrule
\multirow{2}{*}{AF3}
& Self-Con    & \textbf{27.0}  & \textbf{23.7} & \textbf{39.4} & \textbf{47.5} & \textbf{35.3} \\
& Vanilla & 20.0  & 22.2 & 37.6 & \textbf{47.5} & 33.4 \\

\bottomrule
\end{tabular}
\vspace{-10pt}
}
\end{table}
\section{Results}

\subsection{Can Model Steering Improve Chain-of-thought?}
Table~\ref{tab:main_result} presents the results. Steering improves over CoT in most settings, with 11 of 12 model–method combinations showing positive gains in average accuracy. AF3 and Voxtral achieve the largest improvements (+4.4\% and +4.3\%), while other models also benefit from multiple steering variants. Overall, these results suggest that representation interventions can systematically alter CoT outcomes, with effects varying across models.

We compare Vanilla Steering with self-consistency under a comparable computational setup. While both methods use three forward passes, self-consistency performs three full generation processes per instance, whereas Vanilla Steering requires only a single generation pass after extraction. As shown in Table~\ref{tab:main_result_sc}, Vanilla Steering achieves higher overall accuracy on three of the four models. These results indicate that, under a matched forward-pass budget, steering typically attains better accuracy while requiring fewer generation processes.

Beyond Vanilla Steering, we examine two variants, SGS and TGS, which extract general steering vectors without instance-specific information. Both methods improve average accuracy on all models and, in several cases, match or even outperform Vanilla Steering (e.g., TGS on Phi-4-mm and AF3, and both variants on Qwen2.5). These results indicate that general steering directions can transfer across instances while maintaining positive effects on CoT accuracy, providing a practical alternative that avoids instance-specific extraction at test time.


In terms of average performance gain $\Delta$ over the CoT baseline across models, Vanilla Steering achieves an improvement of 1.9\%, SGS 1.4\%, and TGS 2.5\%, with TGS yielding the largest gain. Notably, TGS derives its steering vectors purely from textual data, without using spoken inputs during extraction. This suggests that certain reasoning-relevant representation directions present in the text modality can transfer to spoken tasks. Such cross-modal transfer implies that steering directions may capture modality-agnostic reasoning patterns, offering a lightweight way to enhance spoken reasoning without additional speech-specific training.

\begin{figure}[t!]
    \centering
    \begin{subfigure}{\linewidth}
        \centering
        \includegraphics[width=0.85\linewidth]{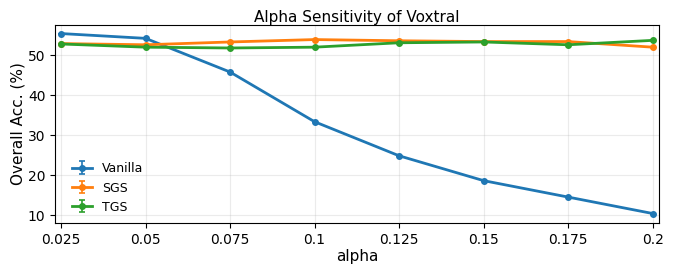}
        \caption{Effect of the scaling factor $\alpha$.}
        \label{fig:alpha}
    \end{subfigure}

    \vspace{1em} 

    \begin{subfigure}{\linewidth}
        \centering
        \includegraphics[width=0.85\linewidth]{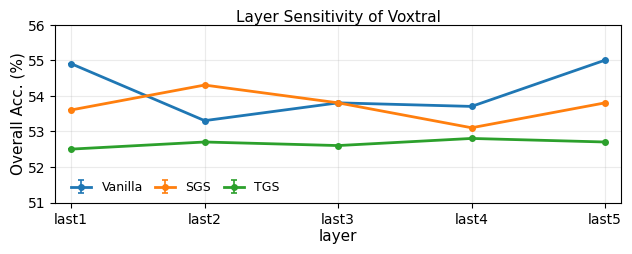}
        \caption{Effect of the number of steered last $k$ layers.}
        \label{fig:layer}
    \end{subfigure}
    \caption{Hyperparameter sensitivity of the steering methods.} 
    \label{fig:hyper}
\end{figure}
\subsection{Hyperparameter Sensitivity of Steering Methods}
\label{sec:hyperparameter}

We analyze the hyperparameter sensitivity of the steering methods on Voxtral, where all methods show clear improvements. Following Table~\ref{tab:main_result}, we vary the scaling factor $\alpha$ while fixing $k$, and vice versa, reporting average accuracy across benchmarks.

As shown in Figure~\ref{fig:alpha}, Vanilla Steering is highly sensitive to $\alpha$: performance peaks at small values and degrades rapidly as $\alpha$ increases. Instance-specific steering vectors capture input-dependent representation shifts, which can be over-amplified under large scaling factors, leading to unstable predictions. In contrast, SGS and TGS remain comparatively stable across a wider range of $\alpha$, likely because their aggregated steering directions induce smoother representation shifts.

Figure~\ref{fig:layer} shows that performance varies non-monotonically with respect to the number of steered last layers. Increasing $k$ does not consistently improve performance, and Vanilla Steering exhibits larger fluctuations than others. These results indicate that steering effectiveness depends on both scaling strength and layer position, with instance-specific steering being more sensitive to hyperparameter choices. Developing automatic strategies for selecting hyperparameters is a promising direction for future research.



\subsection{Data Efficiency of SGS and TGS}

\begin{figure}[t]
  \centering
  \includegraphics[width=0.85\linewidth]{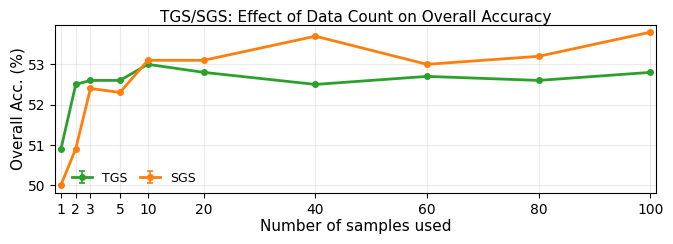}
  \caption{Effect of external dataset size for SGS/TGS on Voxtral.}
  \label{fig:dataset_size}
  \vspace{-10pt}
\end{figure}

We analyze the data efficiency of SGS and TGS by varying $\mathcal{D}^{s}_{\text{ext}}$ and $\mathcal{D}^{t}_{\text{ext}}$ (five-run average; Figure~\ref{fig:dataset_size}).

For SGS, accuracy increases steadily with more spoken samples. Gains are largest at small scales (1–10 samples) and remain evident up to around 40 samples, after which performance largely saturates with diminishing returns. This suggests that estimating a reliable shared steering direction from speech requires a moderate amount of spoken data.

In contrast, TGS remains relatively stable across data scales and reaches near-peak performance even with few textual samples (e.g., 10), indicating robustness to dataset size. Because TGS does not rely on spoken data during extraction, it is more data-efficient and practical when spoken data is limited.

We attribute this stability to the generally stronger and more consistent reasoning performance of LALMs in text-based settings~\cite{sakura, li2025silence}. Since spoken reasoning is typically more challenging~\cite{voxeval, speechr, hsiao25_interspeech}, steering directions derived from text can be estimated more reliably under limited data.



\section{Conclusion}

In this paper, we investigate inference-time model steering as a training-free approach to enhance Chain-of-Thought reasoning in large audio-language models. Across four LALMs and four spoken reasoning benchmarks, we show that representation-level intervention can consistently improve CoT performance. While instance-specific steering achieves strong gains, it is sensitive to hyperparameters and less stable. In contrast, generalized steering directions can be estimated from a moderate amount of auxiliary data and reused across inputs. Notably, text-derived steering provides stable improvements on speech-based tasks with only a few samples, highlighting cross-modal transferability and data efficiency. Overall, our results demonstrate the practical feasibility of model steering for inference-time CoT enhancement in LALMs.

\section{Acknowledgement}
We acknowledge the computational and storage support provided by the National Center for High-performance Computing (NCHC) of the National Applied Research Laboratories (NARLabs) in Taiwan.

\section{Generative AI Use Disclosure}

In this work, generative AI was used only to improve clarity and writing, without contributing to the research content.

\bibliographystyle{IEEEtran}
\bibliography{mybib}

\end{document}